\documentclass[letterpaper,titlepage,11pt]{article}
\usepackage{hyperref}
\usepackage{epsfig}
\usepackage{graphicx}
\usepackage{amsmath}
\usepackage{amsfonts}
\usepackage{amssymb}
\usepackage{mathrsfs}
\usepackage{fancybox}
\usepackage{ragged2e}
\usepackage{tikz}
\usepackage{wasysym}
\long\def\symbolfootnote[#1]#2{\begingroup%
\def\thefootnote{\fnsymbol{footnote}}\footnote[#1]{#2}\endgroup}

\newcommand{\mycite}[1]{~{\cite{#1}}}

\newcommand{\myref}[1]{~{(\ref{#1})}}

\newcommand{\beq}{\begin{equation}}
\newcommand{\eeq}{\end{equation}}

\newcommand{\be}{\begin{equation*}}
\newcommand{\ee}{\end{equation*}}
\newcommand{\alphabar}{\protect{\bar{\alpha}}}
\newcommand{\betabar}{\protect{\bar{\beta}}}
\newcommand{\gammabar}{\protect{\bar{\gamma}}}
\newcommand{\ur}{\mathrm{u}}

\begin{document}

\begin{titlepage}

\rightline{\vbox{\small\hbox{\tt } }}
 \vskip 1.8 cm

\centerline{\LARGE \bf The SO(6) Scalar Product and Three-Point
Functions} \vskip 0.3cm \centerline{\LARGE \bf from Integrability}
\vskip 1.cm

\centerline{\large  {\bf  Agnese Bissi$\,^{1,2}$}, {\bf Gianluca
Grignani$\,^{3}$},  {\bf A.~V.~Zayakin$\,^{3,4}$}}

\vskip 1.0cm

\begin{center}
\sl $^1$ Niels Bohr Institute, Blegdamsvej 17, DK-2100 Copenhagen, Denmark \\
\sl$^2$Niels Bohr International Academy, Niels Bohr Institute, Blegdamsvej 17, DK-2100
Copenhagen, Denmark\\
\vskip 0.4cm
\sl $^3$ Dipartimento di Fisica, Universit\`a di Perugia,\\
I.N.F.N. Sezione di Perugia,\\
Via Pascoli, I-06123 Perugia, Italy
\vskip 0.4cm
\sl $^4$ Institute of Theoretical and Experimental Physics,\\
B.~Cheremushkinskaya ul. 25, 117259 Moscow, Russia
\end{center}
\vskip 0.6cm

\centerline{\small\tt bissi@nbi.ku.dk, grignani@pg.infn.it,
a.zayakin@gmail.com }

\vskip 1.3cm \centerline{\bf Abstract} \vskip 0.2cm \noindent In
1012.2475 Escobedo, Gromov, Sever and Vieira suggested a formula
for an $SU(2)$ three-point correlation function at weak coupling
based on integrability techniques. We conjecture a generalization
of it to the $SO(6)$ sector, thus including all possible
single-trace scalar operators in $\mathcal{N}=4$ super
Yang--Mills, and prove, by direct comparison to a perturbative
$SO(6)$ calculations, that our generalization is valid.

\end{titlepage}

\addtocontents{toc}{\protect\setcounter{tocdepth}{1}}

\setcounter{page}{1}
\section{Introduction}
Three-point functions of single-trace operators in $\mathcal{N}=4$
supersymmetric Yang-Mills have become a foremost research topic
during the last couple of years. The main techniques for
calculating three-point functions may be classified into four
domains: direct perturbative calculation, integrability,
semiclassics and string field theory. Perturbative calculations
and integrability are realized for small 't Hooft coupling
$\lambda$, semiclassics and string field theory for large
$\lambda$. It is generally believed and has been proved
``experimentally'' for some special
cases\mycite{Escobedo:2011xw,Bissi:2011ha,
Grignani:2012yu,Grignani:2012ur}  that the Frolov--Tseytlin
limit\mycite{Frolov:2003xy} gives the regime where weak and strong
coupling results may be compared against each other.

Perturbative calculations were historically the earliest technique
to be used\mycite{Kristjansen:2002bb,Beisert:2002bb} for computing
three-point correlators of non-BPS operators. The direct
perturbative calculations for three-point functions in the spirit
of\mycite{Kristjansen:2002bb,Beisert:2002bb} were performed for
one $\frac{1}{2}$ BPS and two BMN $SU(2)$ states
in\mycite{Okuyama:2004bd}, for two BPS and one BMN $SU(2)$ states
in\mycite{Georgiou:2009tp}, for two $SL(2)$ BMN states and one BPS
state in\mycite{Georgiou:2011qk}, for two BPS states and one
twist-two state\mycite{Plefka:2012rd}, for three short $SO(6)$
operators at one loop in\mycite{Grossardt:2010xq,
Georgiou:2012zj}, for three long $SO(6)$ operators up to one loop
in\mycite{Grignani:2012yu,Grignani:2012ur}. The advantage of this
technique is its straightforwardness, yet dealing with states with
more than two magnons is rather cumbersome. It should be also
noted that perturbative treatment at higher loops is burdened by
extra problems of mixing between bosonic and bifermionic
operators\mycite{Georgiou:2009tp}, mixing between BMN operators of
different momentum\mycite{Beisert:2004ry} and non-trivial
fudge-factors in the wave function\mycite{Staudacher:2004tk}. For
special kinematic cases (extremal correlators) the $1/N$
mixing\mycite{Beisert:2002bb} starts playing its role as well,
already at tree level.

A systematic combinatorial improvement of the perturbative
technique has been developed by means of
integrability\mycite{Escobedo:2010xs}. The general idea of the
method is to operate in the basis of Bethe states rather than in
the field-theoretical basis. Although this application of
integrability techniques that has revolutionized the three-point
function calculations is quite new, the  ingredients needed to
construct the correlators in this way have a long story. The most
non-trivial part of the Bethe Ansatz calculation of a three-point
function is the scalar product of two arbitrary Bethe states. In
1982 Korepin gave a recursive relation for a scalar product of an
$SU(2)$ Bethe eigenstate with an arbitrary $SU(2)$ Bethe
state\mycite{Korepin:1982gg}. This expression was represented in a
concise explicit fashion by Nikita Slavnov in\mycite{Slavnov1989}.
Later the Korepin-Slavnov scalar products, represented as domain
wall partition functions and determinant expressions, were applied
to tree-level three-point correlators in the $SU(2)$ sector
in\mycite{Foda:2011rr,Wheeler:2012bu,Foda:2012yg,Foda:2012wf}. The
scalar products of Bethe states were used to calculate OPE of
non-supersymmetric QCD field-strength operators
in\mycite{Ahn:2012uv}. An integral representation of the scalar
product for a very general class of Bethe Ans\"atze was obtained
in\mycite{deGier:2011rc}. The scalar product of the
Korepin-Slavnov type was derived for the $SU(3)$ generalizations
of the Bethe Ansatz\mycite{resh,Belliard:2012is,Belliard:2012pr}.

The integrability-assisted combinatorics for three-point functions
was originally suggested for tree-level three multi-BMN $SU(2)$
states by Escobedo, Gromov, Sever and
Vieira\mycite{Escobedo:2010xs} (EGSV) at weak coupling.  It was
extended to the semiclassic domain where two of the operators are
heavy and described by semiclassical string states, whereas the
remaining operator is a light mode\mycite{Escobedo:2011xw}, and
also to the light-light-heavy case\mycite{Gromov:2011jh}. Quite
recently the one-loop three-point correlators of multi-BMN $SU(2)$
states were described in terms of the Bethe Ansatz
in\mycite{Gromov:2012uv}. Similar techniques were developed
subsequently by Kostov in\mycite{Kostov:2012jr}, where the
semiclassical result of\mycite{Escobedo:2011xw} was generalized
beyond the semiclassical approximation to an exact determinant
expression for three non-BPS states. In\mycite{Kostov:2012yq} a
factorized operator expression for a scalar product of two
multimagnon states was derived. Another modification of this limit
was suggested by Serban\mycite{Serban:2012dr}.

A lot of interest in understanding the three-point functions from
the point of view of integrability has recently been shown. Kostov
and Matsuo\mycite{Kostov:2012wv} have demonstrated that the inner
product of an on-shell state with $N$ Bethe roots with a generic
$N$-root state is equivalent to a scalar product of the $2N$-root
state with a vacuum descendant state. Three-point functions for
GKP states were found from semiclassical integrability algebraic
curve technique in\mycite{Kazama:2012is}.

In this paper we propose a conjecture for the $SO(6)$ sector
scalar products, thus generalizing the formulation of the three
point function of\mycite{Escobedo:2010xs}. We then calculate the
three-point correlation function for three states which cannot be
embedded into smaller sectors ($SU(2)$ or $SU(3)$) and show that
this structure constant is identical to the one previously found
from string field theory and perturbation theory
independently\mycite{Grignani:2012yu}.

\section{The $SO(6)$ Conjecture}
The starting point for our discussion is  eq. (1.5) for the
three-point function and eq. (A.2) for the scalar product
in\mycite{Escobedo:2010xs}. We generalize these three-point
function and the scalar product to the $SO(6)$ sector and
calculate the correlator of the same states that are used
in\mycite{Grignani:2012yu,Grignani:2012ur}, to enable comparison
between these results. This is yet another step towards the
construction of a general formula for the three point function
valid for any set of local gauge invariant operator in any sector
at weak coupling. For the $SU(2)$ case the expression for the
scalar product has been found
in~\cite{Escobedo:2010xs,Foda:2011rr}. The novelty of these
techniques  have required direct analytic and numeric tests to
ensure that the formula has been reproduced and interpreted
correctly. The $SO(6)$ generalization of the $SU(2)$ formula that
we propose in this paper is a sensible conjecture and it also
requires some tests \footnote{Recently doubts have been raised,
for example, as whether a determinant formula for a scalar product
of two Bethe states can be written down for the $SU(3)$
sector\mycite{Belliard:2012is,Belliard:2012pr}.}. Therefore, we
then specialize our conjecture to the explicit case of three BMN
operators with two impurities, since in this case one can check
the $SO(6)$ formula analytically thanks to the results
of\mycite{Grignani:2012yu}.

Consider a set of operators $\mathcal{O}_A$ normalized to unity
\beq \langle \mathcal{O}_A(x)\label{norm} \bar
{\mathcal{O}}_A(0)\rangle=\frac{1}{x^{2\Delta_A}}. \eeq
The space-time dependence of any three-point function of local gauge invariant operators is
prescribed by conformal symmetry to be
\beq \label{C123}\langle \mathcal{O}_1(x_1) \mathcal{O}_2(x_2)
\bar{\mathcal{O}}_3(x_3)\rangle
=\frac{C_{123}}{|x_{12}|^{\Delta_1+\Delta_2-\Delta_3}
|x_{23}|^{\Delta_2+\Delta_3
-\Delta_1}|x_{31}|^{\Delta_3+\Delta_1-\Delta_2}}. \eeq

The procedure proposed by EGSV\mycite{Escobedo:2010xs} to determine $C_{123}$ in \eqref{C123} is the
following:
\begin{enumerate}
\item Represent each of the operators as a Bethe vector.
\item
Split the rapidities of the three Bethe vectors in all possible
partitions ``cutting'' each of the Bethe vectors into two subvectors.
\item
For each of the partitions calculate the three scalar products of
the states corresponding to the subvectors. Note that in order to have well defined scalar products, one of the two subvectors belonging to each operator has to be ``flipped'', which means that one of the bra is mapped in to a ket by reversing all the spin chain sites (leaving the same charges as the original substate).
\item Sum over all the partitions, taking into account
phase factors due to cutting of the states and conjugating half of
them.
\item Normalize the results to comply with\myref{norm}.
\end{enumerate}
The general expression for tree level and planar structure constant arising from the EGSV procedure then is
\beq \label{partitions}N_c C_{123}=\sum_{\mbox{Root
partitions}}\mbox{Cut } \times \mbox{Flip}\times \mbox{Norm}
\times \mbox{Scalar products}~.\eeq
The structure of the $SO(6)$ formula is conjectured by us to be
analogous to \eqref{partitions} and directly generalizable from it
apart from two subtle issues: the norms and the scalar products.
In the $SO(6)$ sector the Bethe ansatz equations take this form
\begin{equation}
\label{BE}
\left(\frac{\ur_j-i V_{a_j}/2}{\ur_j+i
V_{a_j}/2}\right)^L=\prod_{\stackrel {k=1}{k\neq
j}}^K\frac{\ur_j-\ur_k-\frac{i}{2}M_{a_ja_k}}{\ur_j-
\ur_k+\frac{i}{2}M_{a_ja_k}}
\end{equation}
where $L$ is the length of the chain, $K$ the
number of roots, $M$ is the Cartan matrix and V are the Dynkin labels of the spin representation.
For each of the Bethe roots $\ur_j$ one needs to specify which of the simple roots is excited by $a_j$, which is the number of simple roots and runs from 1 to 3 for the $SO(6)$ sector.
The Cartan matrix for the $SO(6)$ sector is
\beq M_{ab}=\left(\begin{array}{rrr} 2&-1&0\\ -1&2&-1\\ 0&-1&2\\
\end{array}\right),\eeq
and the Cartan weights are \beq V_a=(0,1,0).\eeq
We can see that now we have three sectors instead of one as in the $SU(2)$ sector, thus we represent the magnon $u_i$ by a vector containing its rapidity $\mathrm{u}_i$ and its
level index $a_i$:
\beq u_i=\{\mathrm{u}_i,a_i\} \eeq

Let us consider each of the four elements of the procedure,
cutting, flipping, taking a scalar product and dividing over the
norm of the state, separately. The general idea of EGSV has been
to use integrability techniques and get an analytic expression
for\myref{partitions} in terms of the rapidities of the three
operators. It is well known since~\cite{deVega:1986xj} that the
$SO(6)$ spin chain is integrable, so starting from the R-matrix, which acts on the tensor product of the physical spin chain site vector space and the auxiliary space, and
using its properties, as Yang Baxter algebra, crossing symmetry
and unitarity, it is possible to build a monodromy and a transfer
matrix~\cite{Resh91}. The off diagonal terms of the monodromy matrix can be used
as lowering and raising operators in the sense that acting with
them on a reference state, which is a state with all spins up or
down, one obtains all the possible states. In this context, the Bethe
equations\myref{BE} play a central role ensuring that the states
that we obtain are eigenstates of the transfer matrix. As
explained in details in\mycite{Escobedo:2010xs} the building
blocks used to express the final result of\myref{partitions} are
the functions $f(u)$ and  $g(u)$ (or combination of them) which
appear in the commutation relations of the elements of the
monodromy matrix (see Table 1 of\mycite{Escobedo:2010xs} for the
complete algebra) as well as $a(u)$ and $d(u)$ which instead can
be read from the action on a reference state of the diagonal
elements of the monodromy matrix. In the following, based on these
observations, we propose a conjecture to extend to the $SO(6)$
sector the result for the  three point function of EGSV. The most
natural generalization of the EGSV formula to a general group with
Cartan matrix $M_{ab}$ would follow from replacing the factors
$f,g$ in their expressions (2.29-2.33) by their analogs in higher sectors.
Thus we postulate:
\beq
\begin{array}{l}\displaystyle
f(u_i,u_j)=1+\frac{i M_{a_i a_j}}{2(\ur_i-\ur_j)},\\ \displaystyle
g(u_i,u_j)=\frac{i M_{a_i a_j}}{2(\ur_i-\ur_j)}.\\
\end{array}
\eeq
The indices $a_i,a_j$ are exactly the level indices of the
$i^{\text{th}}$ magnon just defined above.
The $S$-matrix everywhere remains defined as
\beq S(u,v)=\frac{f(u,v)}{f(v,u)}. \eeq
The holonomy factors $a(u),d(u)$ retain their standard definitions
for higher levels
\beq \begin{array}{l} \displaystyle a(u_j)=\ur_j+i V_{a_j}/2,\\
\displaystyle d(u_j)=\ur_j-i V_{a_j}/2,\\ \displaystyle
e(u)=\frac{a(u)}{d(u)}
\end{array}\eeq
so that the Bethe equations have the form \eqref{BE}.
Following EGSV we introduce useful
shorthand notation for products of functions: for an arbitrary
function $F(u,v)$ of two variables and for arbitrary sets
$\alpha,\bar{\alpha}$ of lengths $K,\bar{K}$,
$\alpha=\{\alpha_i\}_K$, $\alphabar=\{ {\alphabar}_{i}
\}_{\bar{K}}$

\beq\begin{array}{l} \displaystyle F^{\alpha,\bar{\alpha}}=
 \prod_{i,j}F(\alpha_i,\bar{\alpha}_j),\\ \displaystyle
F^{\alpha,\alpha}_<=
 \prod_{i<j}F(\alpha_i,\alpha_j),\\ \displaystyle
F^{\alpha,\alpha}_>=
 \prod_{i>j}F(\alpha_i,\alpha_j).
\end{array}
\eeq
For functions $G(u)$ of a single variable let us define
\beq\begin{array}{l}
 G^{\alpha}=
 \prod_{j}F(\alpha_j),\\
 G^{\alpha\pm i/2}=
 \prod_{j}F(\alpha_j\pm i/2).\end{array}
\eeq
Let us take three Bethe vectors $u,v,w$ of lengths $L_1,L_2,L_3$,
corresponding to the operators
$\mathcal{O}_1,\mathcal{O}_2,\mathcal{O}_3$ and split each of
them into two pieces so that the rapidities are such that $\alpha \cup
\alphabar=u, \beta \cup \betabar=v,\gamma \cup \gammabar=w$. The
lengths
$L_\alphabar,L_\alpha,L_\betabar,L_\beta,L_\gammabar,L_\gamma$ of
these pieces are uniquely defined by the possible contraction
structures:
\beq
\begin{array}{l}
L_\alpha=L_\betabar=L_1+L_2-L_3,\\
L_\beta=L_\gammabar=L_2+L_3-L_1,\\
L_\gamma=L_\alphabar=L_3+L_1-L_2.
\end{array}
\eeq
The expression\myref{partitions} will look like
\beq \label{partitions1}
\begin{array}{l} \displaystyle
N_c C_{123}=\sum_{\begin{array}{l} \alpha \cup \alphabar=u\\ \beta
\cup \betabar=v\\ \gamma \cup \gammabar=w\end{array}} \sqrt{L_1
L_2 L_3}\,\, \mbox{\tt Cut}(\alpha,\alphabar) \mbox{\tt
Cut}(\beta,\betabar) \mbox{\tt Cut}(\gamma,\gammabar) \times
\mbox{\tt Flip}(\alphabar)\mbox{\tt Flip}(\betabar)\mbox{\tt
Flip}(\gammabar) \times\\ \displaystyle\hphantom{C_{123}=\sum
LLLLLLLLL} \times \frac{1}{\sqrt{\mbox{\tt Norm}(u) \mbox{\tt
Norm}(v) \mbox{\tt Norm}(w)}} \times \langle \alpha \betabar
\rangle \langle \beta \gammabar \rangle \langle \gamma
\alphabar\rangle\ .
\end{array}\eeq
We work in the ``coordinate'' normalization, where the $\mbox{\tt
Cut}(\alpha,\alphabar)$ factor is organized as
\beq \label{cut}\mbox{\tt Cut}(\alpha,\alphabar)=
\left(\frac{a^\alphabar}{d^\alphabar}\right)^{L_1}
\frac{f^{\alpha\alphabar}f^{\alphabar\alphabar}_<
f^{\alpha\alpha}_<}{f_<^{uu}},\eeq
the factors $\mbox{\tt Cut}(\beta,\betabar)$ and $\mbox{\tt
Cut}(\gamma,\gammabar)$ being analogous to \eqref{cut}. The
$a,d,f,g$ factors are all defined in terms of Bethe Ansatz with
higher-level states taken into account as well. In similar terms
the flip factor may now be rewritten as
\beq \mbox{\tt Flip}(\alphabar)=(e^\alphabar)^L_{\alphabar}\,\,
\frac{g^{\alphabar-i/2}}{g^{\alphabar+i/2}}
\frac{f^{\alphabar\alphabar}_>}{f^{\alphabar\alphabar}_<}, \eeq
analogous expressions work for $\mbox{\tt Flip}(\betabar)$
and $\mbox{\tt Flip}(\gammabar)$.
The norm can also be generalized directly from eq. (5.2) in\mycite{Escobedo:2010xs} and in the coordinate normalization we get
\beq \mbox{\tt Norm}(u)=d^u a^u f^{u u}_>f^{u u}_<
\frac{1}{g^{u+i/2}g^{u-i/2}} \det(\partial_j\phi_k)~, \eeq
here $\partial_j=\frac{\partial}{\partial u_j}$ and the phases
are the ratio of the left and right sides of the Bethe equations
\beq e^{i\phi_j}=e(u_j)^{L_u}\prod_{k\neq j} S^{-1}(u_j,u_k), \eeq
with $a,d,S$ satisfying the multi-level definitions above.

The remaining factor necessary to construct the correlator is the scalar
product. Considering the expression for the scalar product of~\cite{Escobedo:2010xs}
\beq\label{gromov} \begin{array}{l}\displaystyle  \langle v| u
\rangle=g^{uu}_< g^{vv}_> \frac{1}{d^u
a^{v^*}g^{u+i/2}g^{v^*-i/2}f_<^{uu}f_>^{v^*v^*}}\times\\ \\
\displaystyle \hphantom{ \langle v| u \rangle=} \times
\sum_{\begin{array}{l}\alpha\cup\alphabar=u\\ \beta \cup
\betabar=v\end{array}} (-1)^{P_\alpha+P_\gamma}(d^\alpha)^{L_v}
(a^\alphabar)^{L_v} (a^\gamma)^{L_v} (d^\gammabar)^{L_v}\times \\
\\  \displaystyle \hphantom{ \langle v| u \rangle=} \times
h^{\alpha\gamma} h^{\gammabar\alpha}h^{\alpha\alphabar}
h^{\gammabar\gamma}\det t^{\alpha\gamma}
t^{\gammabar\alphabar},\end{array} \eeq
where $t(u)=g^2(u)/f(u),$ and trying to extend the definition towards the $SO(6)$ sector
of the factor $h$ as defined in\mycite{Escobedo:2010xs}
\beq\label{h} h(u)=\frac{f(u)}{g(u)}~,\eeq
the result one gets is not well defined.
In fact $h$ does not
have a direct physical meaning unlike $f$ and $g$. A factor $h$ defined as in~\eqref{h} would be
meaningless since it would then contain division by
zero.

To circumvent this problem  we formulate the $SO(6)$ norm
conjecture via the recursive relation proposed in\mycite{Escobedo:2010xs}, eq.(A.5). This expression is completely regular and
is formulated in terms of physically meaningful objects
$f,g,a,d,S$, thus it makes full sense to conjecture that its
validity extends towards a broader sector. The meaning of this formula goes beyond
the original $SU(2)$ and is supposed to cover the full $SO(6)$
\beq\begin{array}{l} \displaystyle \langle v_1 \dots v_N | u_1
\dots u_N \rangle_N =\sum _n b_n \langle v_1 \dots \hat{v}_n \dots
v_N | \hat{u}_1 \dots u_N \rangle_{N-1} -\\ \\ \displaystyle
\hphantom{\langle v_1 \dots v_N | u_1 \dots u_N \rangle_N =} -
\sum_{n<m} c_{n,m} \langle u-1 v_1 \dots \hat{v}_n\dots \hat{v}_m
\dots v_N | \hat{u}_1 \dots u_N \rangle_{N-1}, \end{array} \eeq
where
\beq b_n=\frac{g(u_1,v_n)\left(\prod_{j\neq n}^N f(u_1,v_j)
\prod_{j<n}^N S(v_j,v_n)-\frac{e(u_1)}{e(v_n)}\prod_{j\neq
n}f(v_j,u_1)\prod_{j>n}S(v_n,v_j)\right)}{g(u_1+i/2)g(v_n-i/2)
\prod_{j\neq 1} f(u_1,u_j)},\eeq
and
\beq\begin{array}{l}\displaystyle
c_{n,m}=\frac{e(u_1)g(u_1-i/2)g(u_1,v_n)g(u_1,v_m)\prod_{j\neq
n,m}f(v_j,u_1)}{g(u_1+i/2)g(v_n-i/2)g(v_m-i/2)\prod_{j\neq
1}f(u_1,u_j)}\times \\ \\ \hphantom{c_{n,m}=}\displaystyle  \times
\left(\frac{S(v_m,v_n)}{e(v_n)}\prod_{j>n}S(v_n,v_j)
\prod_{j<m}S(v_j,v_m)+\frac{d(v_m)}{a(v_n)}\prod_{j>m}S(v_m,v_j)
\prod_{j<n}S(v_j,v_n)\right).
\end{array}\eeq
This will be our working proposal, which shall be checked in a
specific example in the next section. Whenever $g(u_i)$ is used
here as a function of a single argument, it is meant to be
$g(u_i)=\frac{i M_{a_i,a_i}}{2\ur_i}$.

\section{Test of integrability against perturbation theory}
Let us introduce our states as Bethe states. We shall denote an
$N$-root state as
\beq \langle u | = \{\{\ur_1,l_1\},...\{\ur_N,l_N\}\}\eeq
where $u_i$ denotes the value of the rapidity and $l_i$ the level
of Bethe Ansatz it belongs to. The states corresponding to those
studied in\mycite{Grignani:2012yu,Grignani:2012ur} are
\beq
\begin{array}{lll}\displaystyle
\mathcal{O}_1\sim \langle u | &=&\{\{0,1 \},\{\frac{1}{2}\cot
\frac{\pi n_1}{J_1+2},2\},\{-\frac{1}{2}\cot \frac{\pi
n_1}{J_1+2},2\} \},\\ \\ \displaystyle \mathcal{O}_2\sim \langle v
|&=&\{\{0,3 \},\{\frac{1}{2}\cot \frac{\pi
n_2}{J_2+2},2\},\{-\frac{1}{2}\cot \frac{\pi n_2}{J_2+2},2\} \},\\
\\  \displaystyle \mathcal{O}_3\sim \langle w |
&=&\{\{\frac{1}{2}\cot \frac{\pi n_3}{J+1},2\},\{-\frac{1}{2}\cot
\frac{\pi n_3}{J+1},2\} \}.
\end{array}
\eeq
The lengths of the states are $L_1=J_1+2, L_2=J_2+2, L_3=J+2$. The
lengths of substates (or, alternatively, the number of
contractions between each $i$th and $j$th states) are $L_{12}=1$,
$L_{23}=J_2+1$, $L_{31}=J_1+1$;
following\mycite{Grignani:2012yu,Grignani:2012ur} we introduce the
parameter $r$: $J_1=rJ, J_2=(1-r)J$.

Using the definitions of the $SO(6)$ $a,d,f,g,S$ given above we
find all the necessary factors. Expansion in $1/J$ is presumed
everywhere below. We use one of the possible four choices of the
partitions contributing at the leading order in $1/J$
\beq
\begin{array}{ll}
\alpha =\{\{0,1 \},\{\frac{1}{2}\cot \frac{\pi n_1}{J_1+2},2\}
\},& \alphabar =\{\{-\frac{1}{2}\cot \frac{\pi n_1}{J_1+2},2\} \},
\\ \\
\beta =\{\{-\frac{1}{2}\cot \frac{\pi n_2}{J_2+2},2\} \},&
\betabar =\{\{0,3 \},\{\frac{1}{2}\cot \frac{\pi
n_2}{J_2+2},2\}\},
\\ \\
\gamma =\{\{-\frac{1}{2}\cot \frac{\pi n_3}{J+1},2\} \},&
\gammabar =\{\{\frac{1}{2}\cot \frac{\pi n_3}{J+1},2\}\},
\end{array}
\eeq
The flip and cut factors together are
\beq \mbox{\tt Cut}(\alpha,\alphabar)\mbox{\tt
Cut}(\beta,\betabar)\mbox{\tt Cut}(\gamma,\gammabar)\times
\mbox{\tt Flip}(\alphabar)\mbox{\tt Flip}(\betabar)\mbox{\tt
Flip}(\gammabar)=-1 ~, \eeq
the norms yield
\beq \mbox{\tt Norm}(u) \mbox{\tt Norm}(v) \mbox{\tt Norm}(w) = 4
J^2 n_1^2 n_2^2 \pi^4~,\eeq
and the scalar products read
\beq
 \langle
\alpha \betabar \rangle \langle \beta \gammabar \rangle \langle
\gamma \alphabar\rangle  = \frac{n_1 n_2 \sin^2(\pi n_3
r)}{2(n_1-r n_3)(n_2+(1-r)n_3)}~.\eeq
The other contributing partitions in the leading order are
realized by simple transformations $n_1\to -n_1, n_2\to -n_2$.
There are also partitions that contribute at higher orders in
$1/J$, which we do not list here.

Taking all the pieces together we get
\beq \label{c123} N_c C_{123}=-\frac{n_3^2
J^{1/2}(r(1-r))^{3/2}\sin^2(\pi n_3
r)}{(n_2^2-n_3^2(1-r)^2)(n_1^2-n_3^2 (1-r)^2)}, \eeq
which corresponds exactly to the results
of\mycite{Grignani:2012yu} obtained both from perturbation theory and string field theory.

\section{Discussion}

Equation\myref{c123} constitutes the main result of the paper: the
conjectured $SO(6)$ scalar product works non-trivially and yields
precise agreement with perturbative and string-theoretical
calculations from\mycite{Grignani:2012yu}. This is
a stringent test for the validity of our proposal for the extension
of the EGSV method to a sector with
rank greater than one. Only $SL(2)$ and $SU(1|1)$ sectors
have been realized so far, and even for those no comparison to
direct analytic perturbative calculations have been made. Here we stress
again that such a comparison is crucial, although no
such thing as ``integrability calculation of a three-point
function'' exists so far, since it is likely that while
using a complicated combination of $f$'s, $g$'s and $S$'s some
factors may go astray. Surely the extension of the
EGSV method must go further, to the loop
corrections and towards the whole $SU(2,2|4)$.

Our result is obtained from a recursive relation for the scalar
product, it is an important step towards its realization in a more
compact determinant form as was done for the $SU(2)$ case in
\cite{Escobedo:2010xs,Wheeler:2011xw,Foda:2011rr,Wheeler:2012bu}.
Following the derivation in~\cite{Wheeler:2011xw,Foda:2011rr} most likely it
is possible to generalize the scalar products to a
determinant form, but for the norms a regularization prescription
must be provided. Then, it should be possible to write down also
the structure constants in terms of a product of a domain wall
partition function and Slavnov scalar products, which are both
expressed in terms of determinants. This would be extremely
important since it might then lead to a  generalization to all
loops and to any group.

Strong coupling regime must also be tested against integrability
predictions, perhaps with the forthcoming all-loop version. Recent
strong-coupling tests of three-point correlation functions may
seem to have yielded a disagreement for some particular cases.
Namely, it has been shown by Gromov and Vieira
in\mycite{Gromov:2012vu} that semiclassical $SU(2)$ folded string
solutions agree with the $SU(2)$ ``integrability calculations''
only in the strict thermodynamic limit. However this seeming
``disagreement'' does not invalidate any calculations on either
side of the correspondence, since a one-loop weak coupling result
cannot be directly compared to the strongly-coupled result.

 Another crucial example was
provided in\mycite{Bissi:2011ha} where it was shown that the
one-loop correction disagrees with that for semiclassical
calculations of heavy-heavy-light type.  This disagreement may be
due to the approximation of Bethe states with coherent states. A
mismatch between weak and strong coupling results were also
observed in extremal heavy-heavy-light correlators for two giant
gravitons and one point-like graviton~\cite{Bissi:2011dc}. Apart
from giving general arguments on the causes of such (seeming or
true) disagreements, they must be resolved in an exact
quantitative way, and therefore more tests of the ``integrability
three-point functions conjecture'' must be suggested.

\section*{Acknowledgments}

We thank Kolya Gromov for his infinite patience in explaining his
previous work to us. Thanks to Pedro Vieira, Amit Sever and
especially Jo$\tilde{\mathrm{a}}$o Caetano for helping to check
the norm of one of our states. We also thank Omar Foda and
Konstantin Wiegandt for comments on our work. This work was
supported in part by the MIUR-PRIN contract 2009-KHZKRX. The work
of A.Z. is supported in part by the Ministry of Education and
Science of the Russian Federation under contract 14.740.11.0081,
NSh 3349.2012.2, the RFBR grants 10-01-00836 and 10-02-01483.


\providecommand{\href}[2]{#2}\begingroup\raggedright\endgroup

\end{document}